\documentclass[reprint,amsmath,amssymb,aps,twocolumn,showpacs,
superscriptaddress]{revtex4-2}

\usepackage{hyperref}

\usepackage[
]{todonotes}

\definecolor{darkgreen}{HTML}{B2BEB5}
\definecolor{durpurp}{HTML}{742e68}
\definecolor{navy}{HTML}{000080}

\newcommand{\be}{\begin{equation}}
\newcommand{\ee}{\end{equation}}
\newcommand{\ba}{\begin{eqnarray}}
\newcommand{\ea}{\end{eqnarray}}
\newcommand{\beal}{\begin{aligned}}
\newcommand{\eeal}{\end{aligned}}
\newcommand{\beqa}{\begin{eqnarray}}
\newcommand{\eeqa}{\end{eqnarray}}

\DeclareMathOperator\arctanh{arctanh}

\DeclareMathOperator\arcosh{arcosh}

\def\cA{\mathcal{A}}

\begin{document}

\title{Accelerating Black Holes in $2+1$ dimensions: Holography revisited}
\author{Gabriel Arenas--Henriquez}
\email{gabriel.arenas-henriquez@durham.ac.uk}
\affiliation{Centre for Particle Theory, Department of Mathematical Sciences, 
Durham University, South Road, Durham, DH1 3LE, UK}
\author{Adolfo Cisterna}
\email{adolfo.cisterna.r@mail.pucv.cl}
\affiliation{Sede Esmeralda, Universidad de Tarapac\'a, Av. Luis Emilio Recabarren 2477, Iquique, Chile}
\author{Felipe Diaz}
\email{f.diazmartinez@uandresbello.edu}
\affiliation{Departamento de Ciencias F\'isicas, Universidad Andres Bello, Sazi\'e 2212, Santiago de Chile}
\affiliation{Institute of Mathematics of the Czech Academy of Sciences, {\v{Z}}itn\'a 25, 11567 Praha 1, Czech Republic}
\author{Ruth Gregory}
\email{ruth.gregory@kcl.ac.uk}
\affiliation{Theoretical Particle Physics and Cosmology Group, Department of Physics,
King’s College London, University of London,
Strand, London, WC2R 2LS, UK}
\affiliation{Perimeter Institute, 31 Caroline Street North, Waterloo, 
ON, N2L 2Y5, Canada}


\begin{abstract}
This paper studies the holographic description of $2+1-$dimensional accelerating black holes. We start by using an ADM decomposition of the coordinates suitable to identify boundary data. As a consequence, the holographic CFT lies in a fixed curved background which is described by the holographic stress tensor of a perfect fluid. We compute the Euclidean action ensuring that the variational principle is satisfied in the presence of the domain wall. This requires including the Gibbons--Hawking--York term associated with internal boundaries on top of the standard renormalised AdS$_{3}$ action. Finally, we compute the entanglement entropy by firstly mapping the solution to the Rindler--AdS spacetime in which the Ryu--Takayanagi surface is easily identifiable. We found that as the acceleration increases the accessible region of the conformal boundary decreases and also the entanglement entropy, indicating a loss of information in the dual theory due to acceleration.
\end{abstract}
\maketitle

\section{Introduction}
The C-metric, originally found by Levi-Civita \cite{levicmetric} and subsequently by Weyl \cite{weylcmetric}, was first analysed physically by Kinnersley and Walker \cite{PhysRevD.2.1359}, and Bonnor \cite{bonnorcmetric}. It is understood as a pair of causally disconnected black holes that accelerate due to the presence of topological defects, specifically cosmic strings that pull (or struts that push) the black holes away from each other. The spacetime represents a one parameter extension of the Schwarzschild black hole that is subjected to conical defects. The C-metric has been studied extensively  \cite{Letelier:1998rx, Bicak:1999sa, Podolsky:2000at, Pravda:2000zm, Dias:2002mi, Griffiths:2005qp, Krtous:2005ej}, not only in General Relativity (GR) but also in Einstein-dilaton-Maxwell \cite{Dowker:1993bt}, braneworld scenarios \cite{Emparan:1999wa} and in the context of quantum black holes \cite{Emparan:1999fd, Emparan:2000fn, Gregory:2008br, Emparan:2020znc}. Even more, recently supersymmetric extensions were constructed in $D=4$ gauged supergravity \cite{Lu:2014sza,Ferrero:2020twa, Cassani:2021dwa, Ferrero:2021ovq}. These solutions have been uplifted using a Sasaki-Einstein manifold $\rm{SE}_{7}$ to supergravity in $D=11$ resulting in a smooth geometry with properly quantised fluxes. 

The causal structure of the C-metric is fairly well understood, however, its asymptotic structure presents challenges towards a semi-classical analysis of the spacetime. Recently, there has been significant progress in elucidating the thermodynamic behaviour of accelerating black holes \cite{Appels:2016uha, Astorino:2016xiy, Appels:2017xoe, Anabalon:2018ydc, Anabalon:2018qfv, EslamPanah:2019szt, Gregory:2019dtq, Ball:2020vzo, Ball:2021xwt, Gregory:2020mmi, Cassani:2021dwa, Kim:2023ncn, Clement:2023xvq}. A particularly fruitful approach has been where the tensions of the cosmic strings causing the acceleration are considered as thermodynamic variables, which serve as a key ingredient in achieving full cohomogeneity in the first law \cite{Anabalon:2018qfv}.

Another challenging, but less explored, facet of accelerating black holes is the study of acceleration from a holographic point of view. The holographic dual of an accelerating black hole is not yet fully understood, and it remains an active area of research. Some proposals suggest that the dual theory may correspond to a strongly correlated system living in a black hole background \cite{Hubeny:2009kz, Anabalon:2018ydc}. A significant step towards a formal holographic description has been achieved with the discovery of supersymmetric accelerating black holes and their embedding in supergravity and M--theory \cite{Ferrero:2020laf,Ferrero:2020twa,Ferrero:2021etw,Boido:2022iye}, providing a promising avenue for studying these solutions by means of the AdS/CFT correspondence; this approach also seems to suggest the existence of higher-dimensional accelerating solutions, which from a classical geometric perspective have not yet been discovered.

In this direction, an instructive approach is to consider a simple toy model. In this regard, three-dimensional gravity seems to be the perfect candidate to test the boundary properties of accelerating black holes, as the features of two-dimensional field theories are well understood. 
The aim of this paper is to study boundary aspects of the accelerating BTZ black holes, building on the previous investigations of the properties of the space of solutions \cite{Astorino:2011mw,Xu:2011vp, Arenas-Henriquez:2022www}. See \cite{EslamPanah:2022ihg} for the charged case. 

Our paper is organised as follows: In Section II we concisely review the three-dimensional C-metric spacetime and its various classes of solutions. Among the solutions, we focus on the case of an \textit{accelerating} BTZ black hole that is pushed by a strut (negative tension co-dimension one topological defect), as this case exhibits more similarities with the four-dimensional counterpart than the BTZ black hole pulled by a wall (positive tension co-dimension one defect). Note that these solutions include both slow and rapid phases of acceleration, i.e.\ solutions both without and with, respectively, an acceleration horizon in addition to the black hole horizon. Holographically, it can be viewed as simpler to focus on the non-rapidly accelerating solutions so that thermodynamic quantities are uniquely defined.
Next, in Section III, we describe the boundary of the spacetime by employing a radial Arnowitt--Deser--Misner (ADM) foliation. The new ``holographic'' coordinate is aligned with the conformal boundary, such that the boundary metric is easily identifiable. This makes the construction of the stress tensor straightforward. We rewrite the stress tensor using the fluid/gravity correspondence by identifying the pressure and energy density of the dual theory, which lies in a positively curved background. We compute the total energy making use of the holographic stress tensor and analyse the effect of the acceleration. 
Section IV is devoted to computing the Euclidean action, showing that the standard renormalised action for AdS$_{3}$ contains an additional divergence that originates from the domain wall and that extends from the black hole to the boundary. Nevertheless, this divergence is controlled by the inclusion of boundary terms associated with the internal boundaries of the spacetime. The total Euclidean action satisfies the quantum statistical relation upon proper identification of the contribution from the domain wall. 
Section V shows the computation of the holographic entanglement entropy by utilising the relationship between these solutions and Rindler--AdS, finding that the total entanglement decreases with acceleration. 
Finally, we conclude in Section VI with a comprehensive summary and further issues that need to be addressed in the future. Complementary materials are provided in Appendices A and B regarding the explicit details of the FG expansion and of the so-called I$_{\rm C}$ class of accelerating BTZ solutions.

\section{C-metric in 2+1 dimensions: Accelerating black holes}\label{sec:solutions}

We start by describing the three-dimensional C-metric spacetimes presented in \cite{Astorino:2011mw,Xu:2011vp,Arenas-Henriquez:2022www}. A direct truncation of the four-dimensional C-metric allows us to write, in prolate coordinates, the following metric ansatz
\begin{align}\label{eq:metricxy2}
    ds^2 ={}& \frac{1}{\Omega^2}
      \Big[
       -  P(y)d\tau^2
        +\frac{dy^2}{P(y)}
        +\frac{dx^2}{Q(x)}
      \Big]~, \\
      \Omega ={}& A(x-y)~,
\end{align}
where $A$ stands for an acceleration parameter. 
Although these coordinates are not very intuitive, later, we will identify $y$ as a radial coordinate and $x$ as an angular coordinate upon a suitable identification of their ranges. 
The metric polynomials are easily found from the trace of the field equations \cite{Plebanski:1976gy}, which yields two cubic polynomials $P(y)$ and $Q(x)$ of the corresponding coordinates of which, in principle, all polynomial coefficients are arbitrary constants. Imposing the polynomials onto the field equations implies precise relations between the polynomial coefficients. In addition, making use of the symmetries of the line element \eqref{eq:metricxy2}, it is proven that the remaining arbitrariness of the coefficients represents removable gauge redundancies \cite{Arenas-Henriquez:2022www}. In this manner, only sign differences between the polynomial coefficients remain relevant giving rise to three families of accelerating spacetimes, see \autoref{tab:3}.
\begin{table}[!]
\centering
\begin{tabular}{ | c || c | c | c | }
    \hline
    Class   & $Q(x)$    & $P(y)$                  & Maximal range of $x$
    \\
    \hline\hline 
    I       & $1-x^2$   & $\frac{1}{A^2\ell^2}+(y^2-1)$ & $|x|<1$
    \\
    II      & $x^2-1$   & $\frac{1}{A^2\ell^2}+(1-y^2)$ & $x>1$ or $x<-1$
    \\
    III     & $1+x^2$   & $\frac{1}{A^2\ell^2}-(1+y^2)$ & $\mathbb{R}$
    \\
    \hline 
\end{tabular}
\caption[
    Metric functions for the three-dimensional C-metric
  ]{Three different classes of solutions with their maximal range of the transverse coordinate.}
\label{tab:3}
\end{table}
The domain of the $x-$coordinate is chosen such that the metric preserves its signature. 

Generically, Class I represents the geometry of accelerating particle-like solutions, although a particular case dubbed as Class I$_{\rm C}$~, represents an accelerating black hole solution parametrically disconnected from the standard BTZ geometry. As expected from the three-dimensional $\rm{AdS}$ spacetime, it is possible to find ``naked singularities'' (i.e.\ conical solutions corresponding to a ``particle'') in the energy range $-\frac{\pi}{8}\leq M \leq 0$~. Due to acceleration, a Rindler horizon can be formed. For our purposes, besides some computations performed in Appendix \ref{App:Ic} regarding the Class I$_{\rm C}$~, this paper will be devoted to studying the solutions contained in Class II\footnote{Class III does not produce black hole nor particle-like solutions for a single string/wall setting. Instead, it is necessary to introduce a second copy of the spacetime, similar to a Randall-Sundrum scenario \cite{Randall:1999ee, Randall:1999vf}.}. This class is regarded as a one parameter extension of the standard BTZ black hole  \cite{Banados:1992gq,Banados:1992wn} and thus we shall denote it as the accelerating BTZ black hole \cite{Astorino:2011mw, Arenas-Henriquez:2022www}. Taking the parameter $A \rightarrow 0$~, we recover the standard geometry of the one paprameter family of three-dimensional black holes. This is key when comparing the black hole solutions of Class II and Class I$_{\rm C}$~. The latter exists only for a limited range of parameters where the acceleration and mass of the black hole are bounded, and the BTZ geometry is not included in this range, although it has a compact horizon.  

It is worth noting that both Class I and Class II spacetimes possess a well-defined flat limit
\begin{align}
    \lim_{\ell\to \infty}ds^2 ={}& \frac{1}{\Omega^2}\left(-P_\infty (y) d\tau^2 + \frac{dy^2}{P_\infty (y)}+\frac{dx^2}{Q(x)} \right)~,
\end{align}
where $P_\infty (y) = \mp(1-y^2)$~, and where Class I corresponds to a minus sign and Class II for the opposite. Here $\Omega$ denotes the acceleration conformal factor. These solutions are interpreted as accelerating particles moving on a three-dimensional flat background \cite{Anber:2008qu, Anber:2008zz}. 

In the next subsection, we summarise the main details behind the construction and interpretation of the solutions contained in Class II. As we shall shortly observe, to construct the solutions it is necessary to introduce a domain wall in the spacetime and to use Israel junction conditions to identify the tension (positive or negative) of the wall, which is responsible for the ``acceleration'' \cite{Arenas-Henriquez:2022www}. The wall extends from the horizon to the boundary, therefore affecting the definition of holographic quantities as we will explicitly demonstrate.

\vspace{5mm}
\paragraph{\textbf{Inserting a domain wall}.}

\begin{figure*}
\centering
\includegraphics[width=\textwidth]{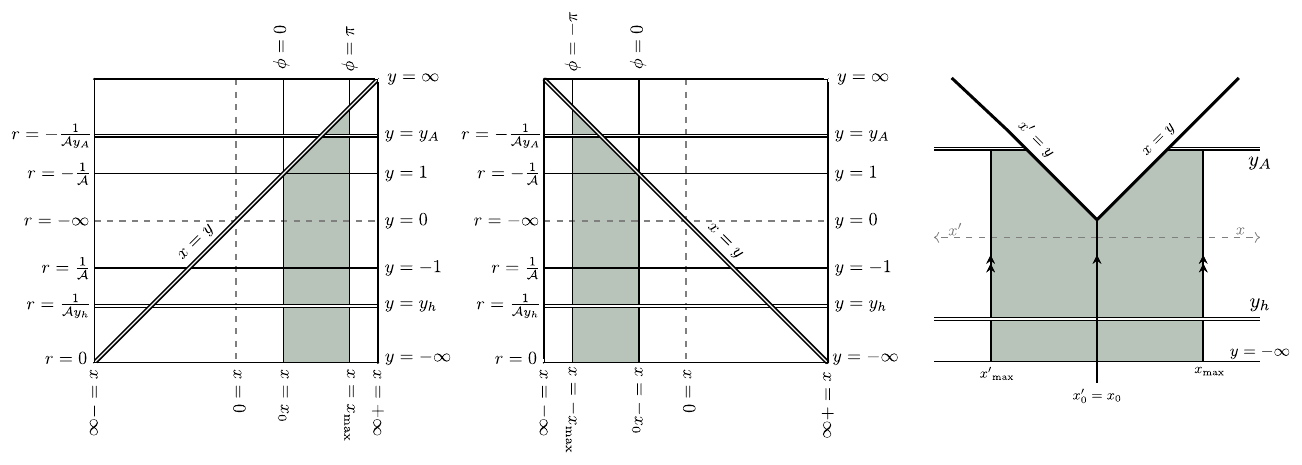}
\includegraphics[width=\textwidth]{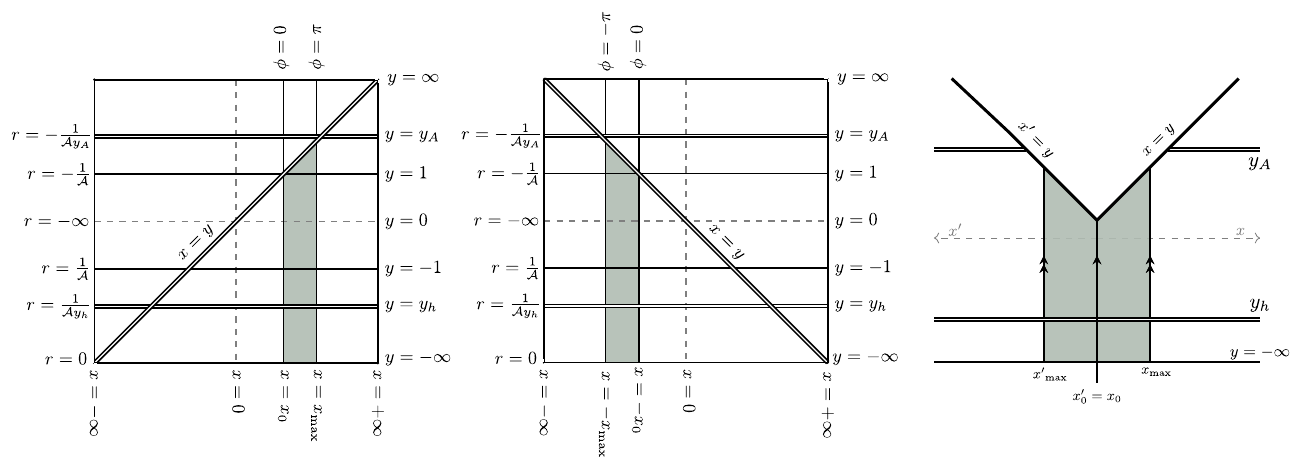}
\includegraphics[width=\textwidth]{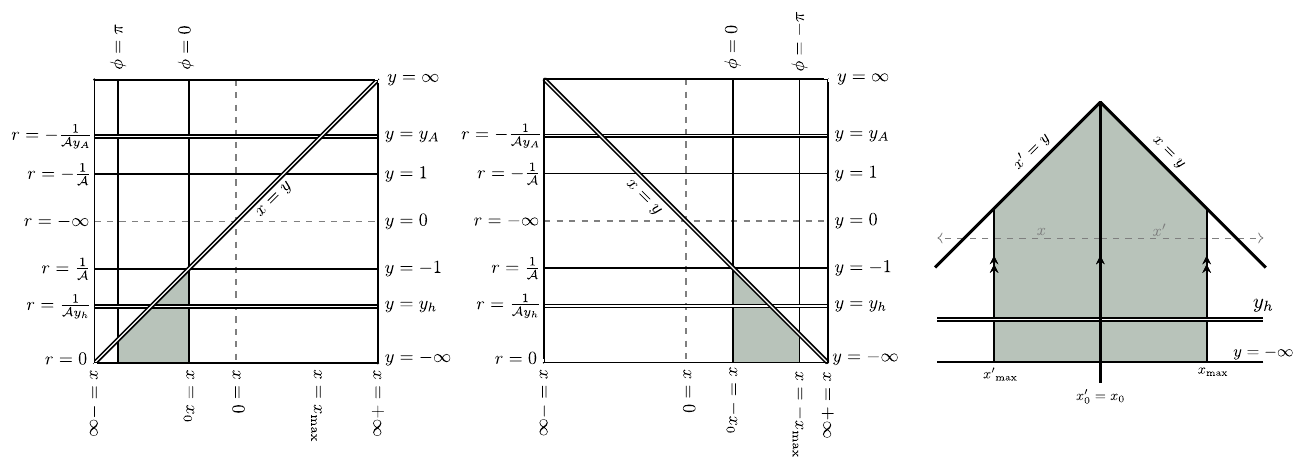}
 \caption{Here, we provide the schematic construction of the three different kinds of black hole solutions contained in class II. They are ordered as follows: Row I shows the construction of the accelerating BTZ black hole pushed by a strut for which the $x-$coordinates satisfies $x>1$ and where both event and accelerating horizons are part of the spacetime causal structure. The second row expresses the construction of the same previous solution, but in the case in which the accelerating horizon is absent. 
Finally, the third row shows the construction of the accelerating BTZ black hole pulled by a domain wall, solution for which $x<-1$~. All diagrams correspond to a constant time coordinate. 
The first picture of each row represents the preliminary causal structure, in prolate and polar coordinates, of the given solutions. This allows us to understand the range of the coordinates and the position of the corresponding horizons. The second pictures correspond with the mirroring of the first ones, while the third represents the final form of the spacetime once the gluing has been performed.
}
\label{fig:c2const}
\end{figure*}

In dimension four a straightforward inspection of the axes of symmetry of the C-metric line element reveals the existence of conical singularities. It is the difference in deficit angle of the conical singularity at both the north and south poles that gives rise to the acceleration. This conical singularity can be either an angular deficit or excess, but the singularity can be removed along an axis by choosing the periodicity of the azimuthal coordinate. This transfers the full defect to one polar axis, usually an angular deficit along the south pole. Physically, this is then interpreted as a cosmic string that extends all the way from the horizon to conformal infinity, and that pulls the black hole producing its acceleration \cite{stephani2003,griffiths2009,PhysRevD.34.2263,Gregory:2017ogk}.  

In $(2+1)-$dimensions the situation is remarkably different. The first difference relies on the fact that with one dimension less the topological defect causing acceleration, which is linear in nature, will now have co-dimension one -- i.e.\ will be a domain wall rather than a cosmic string. The second difference relates to the nature of the angular coordinate. In three dimensions the metric functions depend on an azimuthal angle (not polar as in the four-dimensional case) and therefore the domain of this coordinate also behaves differently. It was shown in \cite{Arenas-Henriquez:2022www} that in order to have a compact horizon the $x-$coordinate needs to be properly identified, which is precisely determined by the inclusion of the domain wall that is responsible for the black hole acceleration.  

To construct the accelerating Class II BTZ geometry the following procedure is taken:\\
i) First, we define a finite domain for the $x-$coordinate, $[x_0,x_{\rm max}]$~, where $x_0$ is either greater than $1$ or smaller than $-1$~, see Table (\ref{tab:3}). The value of $x_{\rm max}$ is constrained according to the number of Killing horizons we allow our geometry to contain. Generically, $P(y)$ provides two Killing horizons located at 
\begin{equation}
y_A=\frac{\sqrt{1+A^2\ell^2}}{A\ell}, \hspace{0.3cm}y_h=-\frac{\sqrt{1+A^2\ell^2}}{A\ell}.
\end{equation}
Here, $y_A$ and $y_h$ represent the acceleration and black hole horizons respectively. Notice that $y_A>1$~. The position of $x_{\rm max}$ then determines the number of horizons according to whether $x_{\rm max}<y_A$~, in which case only the black hole horizon is present, or $x_{\rm max}>y_A$~, in which case both horizons are present.\\
ii) After specifying the interval $[x_0,x_{\rm max}]$~, we proceed with the construction of a compact horizon by identifying two copies of the spacetime along two surfaces of constant $x$~, mirroring along $x_0=1$ and $x_{\rm max}$ (we take $x_0=1$ so that the ``wall'' along the mirrored surface has zero tension). This is carried out by including a domain wall at $x_{\rm max}$~, of which the induced line element reads, 
\begin{align}
ds^{2}_{\rm{DW}} ={}& \gamma_{MN}dx^M dx^N \nonumber \\ ={}& \frac{1}{A^2(x_{\rm max}-y)^{2}} \left( -P(y) d\tau^{2} + \frac{dy^{2}}{P(y)} \right)~,\label{branemetric}
\end{align}
where $N,M=(\tau,y)$ are the domain wall coordinates. The wall is sourced by a localised energy-momentum tensor, of which the integration over the thin wall configuration, using Israel equations, provides 
\begin{equation}\label{Israel}
4 \pi G \int_{-}^{+} T_{MN}= \left[\mathcal{K}_{MN}\right]\big|_-^+-\gamma_{MN}\left[\mathcal{K}\right]\big|_-^+= 4\pi G\sigma \gamma_{MN}~,
\end{equation}
where $T_{MN}$ is the energy stress tensor of the domain wall. Here $G$ is the Newton's constant, the brackets $[{\cal K}]\rvert_-^+$ correspond to the difference of the extrinsic curvature along each side of the domain wall, and ${\cal{K}}_{MN}:=\tfrac12 {\cal L}_n \gamma_{MN}$ is the extrinsic curvature of the hypersurface at $x=x_{\rm max}$ which is given by the covariant derivative of the outward pointing normal 
\begin{equation}
n_{x}= \frac{1}{A(x-y) \sqrt{Q}} \frac{\partial}{\partial x}\bigg |_{x_{\rm max}}~, \label{normaln}
\end{equation}
yielding
\begin{equation}
\sigma = -\frac{1}{4\pi G}\left[{\cal K}\right]\rvert_-^+ =\pm \frac{A}{4 \pi G} \sqrt{Q(x_{\rm max})}~.   \label{tension} 
\end{equation}
With these steps at hand, the construction of the Class II solutions is complete. One sector of solutions is constructed from $x>1$~, while the other follows $x<-1$~. In both cases $x-y>0$~. For the $x>1$ case the domain wall has negative tension, contrary to its $x<-1$ cousin for which the domain wall acquires a positive tension. These geometries are dubbed as the accelerating BTZ black hole pushed by a strut and the accelerating BTZ black hole pulled by a wall, respectively. In this paper, we will mostly focus on the spacetime described by $x>1$~. Within these solutions, we note that if $x_{\rm max}>y_A$~, the acceleration horizon is included in our spacetime for the region $x>y_A$ this is the situation referred to as \emph{rapid acceleration}. 
This phenomenon does not occur for the accelerating BTZ pulled by a wall, as the defect on the horizon in this case has a positive energy density. Since the wall pulls the horizon far from the conformal boundary, no rapid accelerating phase takes place. The causal structure of both solutions and their construction by gluing is depicted in figure \ref{fig:c2const}.

\medskip
\paragraph{\textbf{Accelerating BTZ black hole pushed by a strut.}}

We now focus on the accelerating BTZ black hole pushed by a strut. While the prolate coordinates $(x,y)$ are useful for describing the construction and causal structure of the solutions, it is convenient to move to the more intuitive polar coordinates $(r,\phi)$ to describe the holographic properties of the spacetime. 
Just as in the four-dimensional C-metric, the coordinate $y$ is easily identified with the radial polar direction $r$~. Although $x$ is in principle non-compact, due to the procedure to introduce the domain wall and have a proper black hole interpretation it becomes compact and can be related with an angle. 
Thus, we introduce a mass parameter and new coordinates via\footnote{Notice that the parametrisation is such that the limit ${\cal{A}}\to0$ renders the metric identical to the standard BTZ black hole with mass parameter $m^2$~.}
\begin{align}
r = -\frac{1}{{\cal A}y}~, \quad x = \cosh(m\phi)~, \quad t  =\frac{m^2\cA}{\alpha}\tau~,
\end{align}
where $\cA = A/m$~, and $\alpha$ is a constant that will be used to identify the proper time of an asymptotic observer \cite{Anabalon:2018qfv}. The resulting metric reads 
\begin{align}\label{gtrphi}
ds^2 = \frac{1}{\Omega^2}\left(-f(r)\frac{d\tau^2}{\alpha^2} + \frac{dr^2}{f(r)} + r^2d\phi^2\right)~,
\end{align}
where
\begin{align}
f(r) ={}& \frac{r^2}{\ell^2} - m^2(1-\cA^2r^2)~,\nonumber \\ \Omega ={}& 1+\cA r \cosh(m\phi)~.
\end{align} 
The tension of the wall is regulated by the parameter $m=\arcosh(x_{\rm max})/\pi$ that is chosen to ensure that the coordinate $\phi$ lies in the interval $(-\pi,\pi)$~. 
This is consistent with the standard interpretation of mass of three-dimensional black holes \cite{Banados:1992wn}.  Nevertheless, the non-trivial extrinsic curvature over the $x=\rm{const}$ surface modifies the original identifications used to construct the BTZ black hole. 
Moving forward, our main focus will be on describing this specific solution, although all of our results can be applied to the positive tension scenario with ease.

Considering now the Euclidean version of the solution, the horizon radius, Hawking temperature, and entropy are easily read off as  \cite{Arenas-Henriquez:2022www}
\begin{align}\label{TS}
r_h ={}& \frac{m\ell}{\sqrt{1+m^2\cA^2\ell^2}}~,\\ \nonumber T ={}& \frac{|f'(r_h)|}{4\pi\alpha} = \frac{m\sqrt{1+m^2\cA^2\ell^2}}{2\pi\ell\alpha}~,\\ S ={}& \frac{\ell}{G} \arctanh\left[\left( \sqrt{1+m^2\cA^2\ell^2}-m\cA\ell\right)\tanh\left(\frac{m\pi}{2}\right)\right]~,\nonumber 
\end{align}
where ${\rm Area}$ refers to the area of the black hole horizon and the conformal boundary is now given by $r_{\rm cb}=-(\cA \cosh(m\phi))^{-1}$~.
The wall lies along $\phi=\pm\pi$~, with tension 
\begin{align}\label{Tension}
    \sigma = -\frac{ m\cA \sinh(m\pi)}{4\pi G}~.
\end{align}
As already mentioned, there is a rapid phase when the wall position is $x_{\rm max} > \sqrt{1+(m\cA\ell)^{-2}}$~, implying the appearance of a non-compact accelerating horizon \cite{Arenas-Henriquez:2022www}. In order to avoid this feature for the holographic analysis of the solution, one restricts $m$ to be positive and to satisfy the condition 
\begin{align}\label{slowacc}
m\cA \ell \sinh(m\pi) < 1~,  
\end{align}
which is referred to as the condition of slow acceleration. 

Accelerating black holes pulled by a domain wall (positive tension) can be found by changing the sign of the acceleration $\cA \to - \cA$~. In that case, there is no rapid accelerating phase, and therefore the only constraint for the acceleration is given by requiring a positive radial coordinate. See Fig.\ \ref{fig:horzpsi}.
It is also possible to recover the accelerating particle solutions of Class I by taking $m^2 \to -m^2$~, noting that the hyperbolic cosine becomes a cosine in the conformal factor $\Omega$ (the Class I$_c$ solutions require a slightly more subtle transformation). Finally, it is worth mentioning that the solution can be mapped to a three-dimensional Rindler geometry
 \begin{align}\label{Rindler}
    ds^2 = -\left(\frac{R^2}{\ell^2} - 1\right)dT^2+\frac{dR^2}{\frac{R^2}{\ell^2} - 1}+R^2d\Theta^2~,
\end{align}
with $R \in (\ell,\infty)$~, by means of 
\begin{align}\label{Rmap}
    \frac{R^2}{\ell^2} - 1 ={}&\frac{f(r)}{\alpha^2 m^2\Omega^2}~, \quad R\sinh\Theta = \frac{r\sinh(m\phi)}{m\Omega}~,
\end{align}
provided that
\begin{align}
    T = m\tau~, \quad \alpha^2 = 1+m^2\cA^2\ell^2~.
\end{align}
This identifies the Rindler time $T$~and the value of $\alpha$~. 

In the next section, we will introduce an alternative coordinate system that simplifies the identification of the boundary structure, and use it to characterise the holographic stress tensor.

\section{Holographic stress tensor}\label{Sec:HoloTmunu}
\paragraph{\textbf{ADM-like coordinates.}}

A crucial step towards the identification of the  thermodynamic quantities and holographic data is to have at hand a robust description of the spacetime boundary. 
In this regard, the asymptotic structure of the C-metric poses a challenge, as the conformal boundary is not given by a constant value of the radial coordinate. 

The standard recipe, when treating asymptotically AdS spacetimes, is to apply a Fefferman-Graham (FG) expansion near the boundary, identifying the holographic coordinate.  In the case of the C-metric, however, this process is not trivial but rather complicated.  This was first noticed for the four-dimensional accelerating black hole in \cite{Anabalon:2018qfv}, where an asymptotic expansion for both the radial and angular coordinates was derived in terms of the FG holographic coordinate that is perpendicular to the boundary (see Appendix \ref{App:FG} for the procedure in the three-dimensional case). 

Solving Einstein's equations order by order in the expansion reconstructs the spacetime and gives a boundary metric $g_0$~, which is defined up to a conformal representative $\omega$~. In four dimensions, the Euclidean action and conserved quantities (such as the mass) are independent of the conformal representative of the boundary metric. However, in three dimensions, the situation is different. The conformal invariance is broken at the quantum level, and the dual two-dimensional CFT has a conformal anomaly. The anomaly itself is independent of the conformal factor and reproduces the value of the Brown-Henneaux central charge \cite{Brown:1986nw} for any representative Class \cite{Arenas-Henriquez:2022www}. Nevertheless, the choice of $\omega$ is crucial to identify the mass and the other thermodynamic quantities since the dual stress tensor, which is employed to obtain the holographic mass, is not a primary operator and transforms non-trivially under conformal transformations. For a discussion on how the energy, action, and other holographic quantities depend on the conformal representatives, see \cite{Papadimitriou:2005ii}. 

An alternative way of obtaining the boundary data is to follow \cite{Hubeny:2009kz,Cassani:2021dwa} and define a new coordinate $z$ according to 
\begin{align}\label{z}
    \frac{1}{r} = z - \cA \cosh(m\phi)~.
\end{align}
In this new coordinate frame, the location of the conformal boundary is at $z_{\rm cb} = 0$~. We introduce an infrared cutoff at $z = \delta$~, where $\delta<<1$~.

\begin{figure}[t]
\begin{center}  
\includegraphics[scale=0.45]{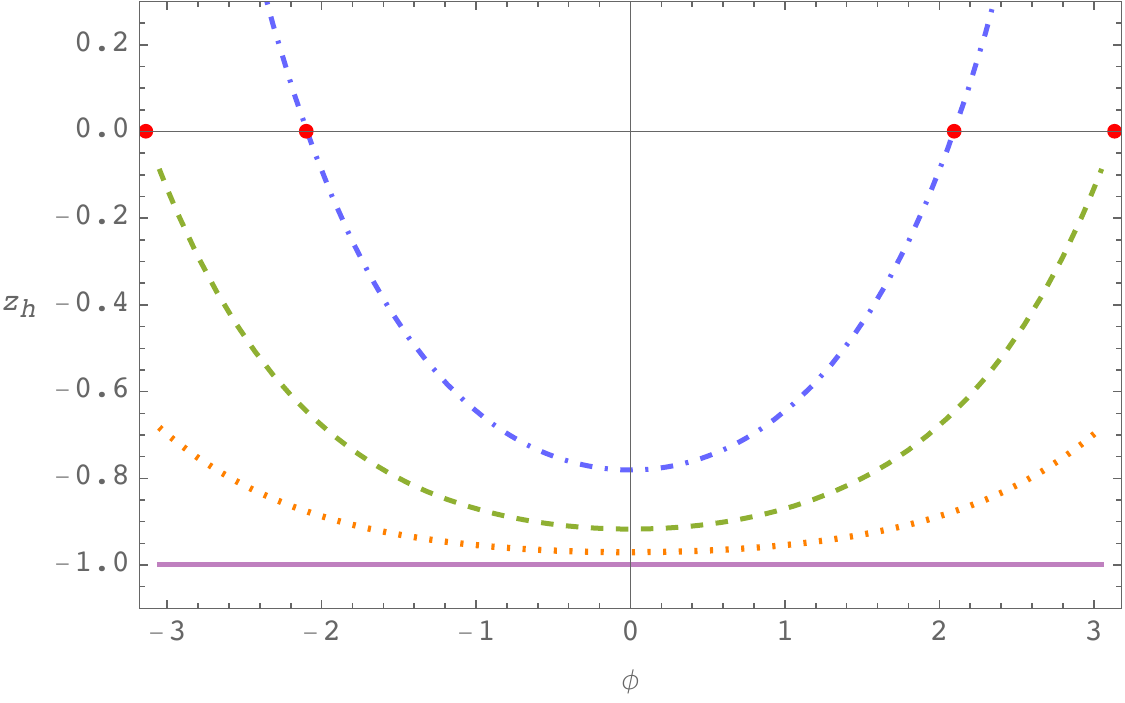} 
\caption{Accelerated horizon radius as a function of the azimuthal coordinate $\phi$~, for different values of the parameter ${\cal A}$ that characterises the acceleration of the black hole. The solid purple line corresponds to the case ${\cal A} = 0$~, where both horizons coincide, and therefore remain at a constant value of $z$~. The orange dotted curve corresponds to ${\cal A} = (2\ell m\sinh m\pi)^{-1}$~, an intermediate value where the horizon remains smooth throughout. The green dashed curve corresponds to the critical value ${\cal A} = (\ell m\sinh m\pi)^{-1}$~, where the horizon touches the conformal boundary exactly at the endpoints $\phi=\pm \pi$~. The blue dot-dashed curve corresponds to ${\cal A} = 3(\ell m\sinh m\pi)^{-1}$~, which is below the critical bound and the red dots indicate the points where the horizon meets the conformal boundary.}
\label{fig:horzpsi}
\end{center}
\end{figure}

\begin{figure}[ht]
\begin{center}  \includegraphics[scale=0.45]{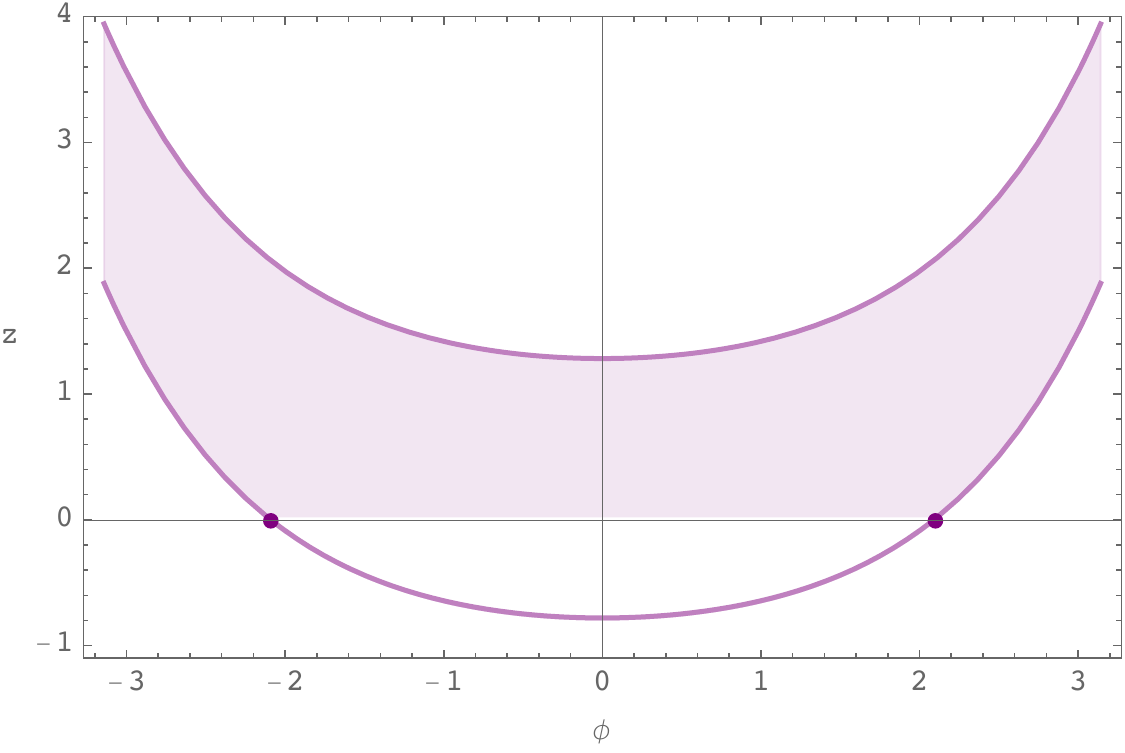} 
\caption{Domain of the $z$ coordinate in the presence of the two horizons. The upper thick curve corresponds to the black hole horizon and the bottom one corresponds to the accelerating horizon touching the conformal boundary at the two dark dots. The shaded region corresponds to the region where spacetime is defined.}
\label{fig:Acchorzpsi}
\end{center}
\end{figure}

Following the standard prescription \cite{deHaro:2000vlm}, we first compute regularised holographic quantities close enough to the boundary at $z = \delta$~, to then take the limit $\delta\to 0$~. In this coordinate system, the horizon is described by a function $z_h = z_h(z,\phi)$~. The accelerated horizon is plotted in \autoref{fig:horzpsi}, and considering both horizons the considered region can be seen in \autoref{fig:Acchorzpsi}. The slow acceleration condition \eqref{slowacc} ensures that the horizon does not touch the conformal boundary $z_{\rm cb}$~.

This coordinate transformation sets the metric into an ADM-like decomposition
\begin{align} \label{adm}
ds^2 = N^2 dz^2 + h_{ij}(dx^i + N^i dz)(dx^j + N^j dz)~,
\end{align}
where $h_{ij}$ is the induced metric in the $z = {\rm const}$ hypersurface, $N$ the lapse function and $N^{i}$ represents the shift vector.
Notice that the induced metric $h_{ij}$ depends on $z$ and $\phi$~. We can identify the background in which the holographic CFT lies to be
\begin{align}\label{g0metric}
ds_{(0)}^2 ={}& g_{(0)ij}dx^idx^j = \lim_{\delta\to 0}\delta^2 h_{ij}dx^idx^j  \nonumber \\ ={}&G(\xi)\left(-d{\tilde \tau}^2 
 + d\xi^2\right)~,
 \end{align}
where the corresponding coordinates 
\begin{align}
    \tau &= \alpha\ell{\tilde{\tau}}~,\quad \xi = \frac{\arctanh\left(\alpha\tanh(m\phi)\right)}{m\alpha}~,
 \end{align}
 and conformal factor
 \begin{align}\label{Gomega}
     G(\xi) &= \frac{2\alpha^2}{2+\alpha^2\left(1+\cosh(2m\xi\alpha)\right)}~.
 \end{align}
The extrinsic curvature associated with the boundary metric $h_{ij}$ reads
\begin{align} \label{extrinsich}
    K_{ij} := \frac12\mathcal{L}_n g_{ij} = -\frac{1}{2N}\left(\partial_{z}h_{ij} - \nabla_i N_j - \nabla_j N_i\right)~,
\end{align}
with $\nabla_i$ the covariant derivative respect to $h_{ij}$~, and where the outward-pointing normal to the $z= \rm{const}$ hypersurface is
\begin{align}
    n = \frac{1}{N}\left(N^i \partial_i-\partial_z \right)~.
\end{align}
Finally, its trace is taken with respect to the boundary metric, $K = h^{ij}K_{ij}$~. An important cross-check of the behaviour of the solution consists of analysing the leading order term of the extrinsic curvature near the boundary
\begin{align}
K^i_{~j} \sim \left(\frac{1}{\ell}+{\cal O}\left(z\right)\right)\delta^i_{~j}~.
\end{align}
The asymptotic behaviour described above represents the minimum requirement in which the gravitational action has a well-posed variational principle, enabling  the definition of holographic conserved quantities in AdS$_3$ \cite{Miskovic:2006tm}. It should be noticed that the coordinate $z$ defined in \eqref{z} matches the FG coordinate only at the leading order, which is sufficient for constructing the holographic stress tensor and the boundary conformal classes $g_{(0)}$~.  However, it is important to highlight that the holographic free energy is not guaranteed to match in this coordinate frame due to the discrepancy between the two gauges. In the presence of an odd-dimensional bulk, the conformal freedom in the FG frame transforms the free energy non-trivially and therefore, the free energy depends on the choice of conformal representative of the boundary theory \cite{Papadimitriou:2005ii}. We explore this issue in \autoref{sec:HoloAction}. 
\vspace{3mm}
\paragraph{\textbf{Boundary stress tensor and holographic energy.}}
The holographic energy-momentum tensor is given by the variation of the regularised action  with respect to the first term of the FG expansion $g_{(0)}$~. This has been written in terms of quantities depending on the induced metric \cite{Balasubramanian:1999re}, which in dimension three yields 
\begin{align}\label{Tmunu}
    \langle T_{ij}\rangle ={}& \lim_{z\to 0} -\frac{1}{8\pi G}\left(K_{ij}-K h_{ij} + \frac{1}{\ell}h_{ij} \right)~.
\end{align}
Using \eqref{gtrphi} we obtain
\begin{align}
    \langle T^\tau_{~\tau}\rangle ={}& -\frac{m^2\ell}{32\pi G}\left(2+m^2\cA^2\ell^2 - 3m^2\cA^2\ell^2\cosh(2m\phi)\right)~, \nonumber \\ \langle T^\phi_{~\phi}\rangle ={}& \frac{m^2\ell}{16\pi G}\left(1+m^2\cA^2\ell^2 \cosh^2(m\phi)\right)~,
\end{align}
which is, indeed, covariantly conserved with respect to the boundary metric $g_{(0)}$~, viz., $\nabla_i^{(0)} T^{ij} = 0$~. In addition, it should be pointed out that the stress tensor exhibits a non-vanishing trace, indicating the breakdown of Weyl invariance in the quantum theory, and resulting in the emergence of the conformal anomaly
\begin{align}\label{weylanomaly}
    \langle T^i_{~i}\rangle = \frac{c}{24\pi}{\cal R}[g_{(0)}]~.
\end{align}
Here, $c = 3\ell/2G$ matches the Brown--Henneaux central charge \cite{Brown:1986nw} and ${\cal R}[g_{(0)}] = 2m^4\cA^2\ell^2\cosh(2m\phi)$ is the curvature of the boundary metric, which is always positive. Additionally, it is worth noting that the stress tensor can be expressed in the form of a perfect fluid
\footnote{In 1+1 dimensions there is no shear nor viscosity and it is still possible to have non-constant pressure while maintaining the form of a perfect fluid. See \cite{Ciambelli:2020eba} for an analysis of two-dimensional holographic fluids.}
\begin{align}\label{fluidT}
\langle T_{ij} \rangle =  \left(p + \rho\right)u_i u_j  + p g_{(0)ij}~,
\end{align}
on a curved background $g_{(0)}$ and with timelike velocity $u^i$ given by
\begin{align}
    u^i = \frac{1}{\sqrt{-g_{(0)tt}}}\left(\frac{\partial}{\partial t}\right)^i~,\quad u_iu_j g_{(0)}^{ij}=-1~.
\end{align}
The energy density $\rho$ and pressure $p$ read
\begin{align}
    \rho ={}& \frac{m^2\ell}{32\pi G}\left\{2+m^2\cA^2\ell^2\left[1-3\cosh(2m\phi)\right]\right\}~, \nonumber \\ 
    p ={}& \frac{m^2\ell}{16\pi G}\left[1+m^2\cA^2\ell^2\cosh^2(m\phi)\right]~.
\end{align}
We observe that the energy density $\rho$ generates the energy flow of the fluid since $u_i T^{ij} = -\rho u^j$~. It is noteworthy that this is different from the dual of the four-dimensional accelerating black hole, where the stress tensor cannot be expressed in the form of a perfect fluid, and corrections due to the acceleration parameter cause the boundary metric to be non-conformally flat, leading to non-trivial stress tensor components \cite{Anabalon:2018ydc, Anabalon:2018qfv}.

\begin{figure}[t]
\begin{center}  \includegraphics[scale=0.45]{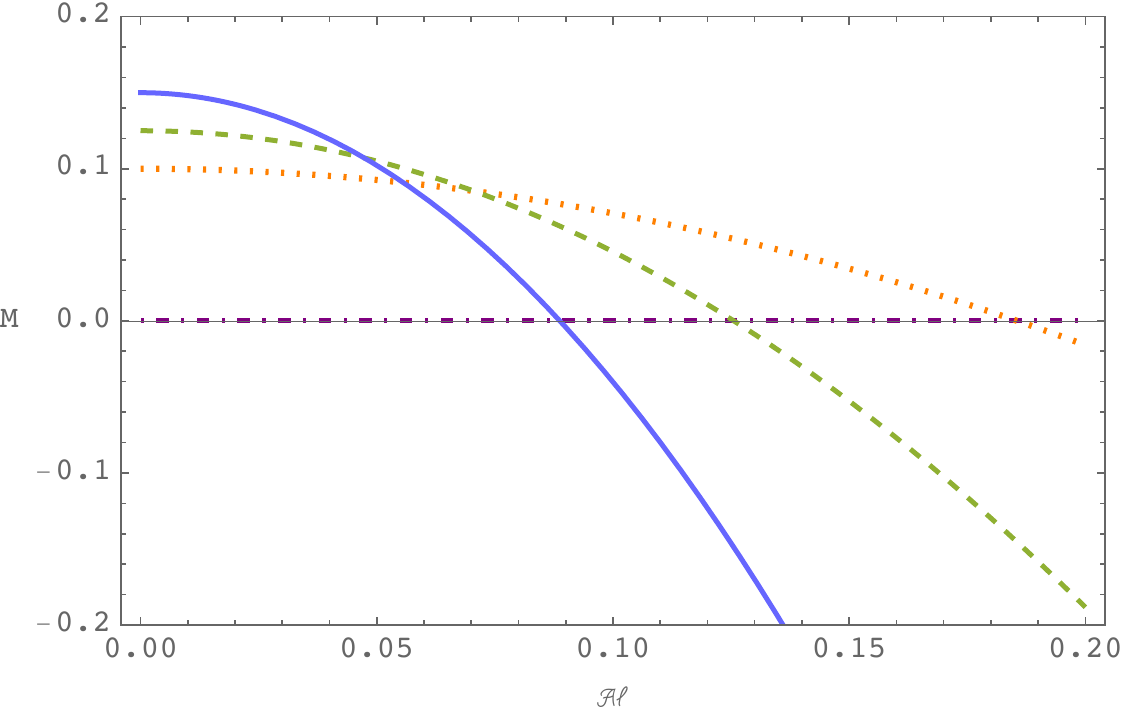} 
  \caption{Holographic mass of the accelerated BTZ pushed by a strut with respect to $\cA \ell$~, with $m^2 = 0,0.8,1,1.2$ for the dot-dashed purple, dotted orange, dashed green, and solid blue curves, respectively.}
  \label{fig:massblA}
  \end{center}
\end{figure}

\begin{figure}[t]
\begin{center}  \includegraphics[scale=0.45]{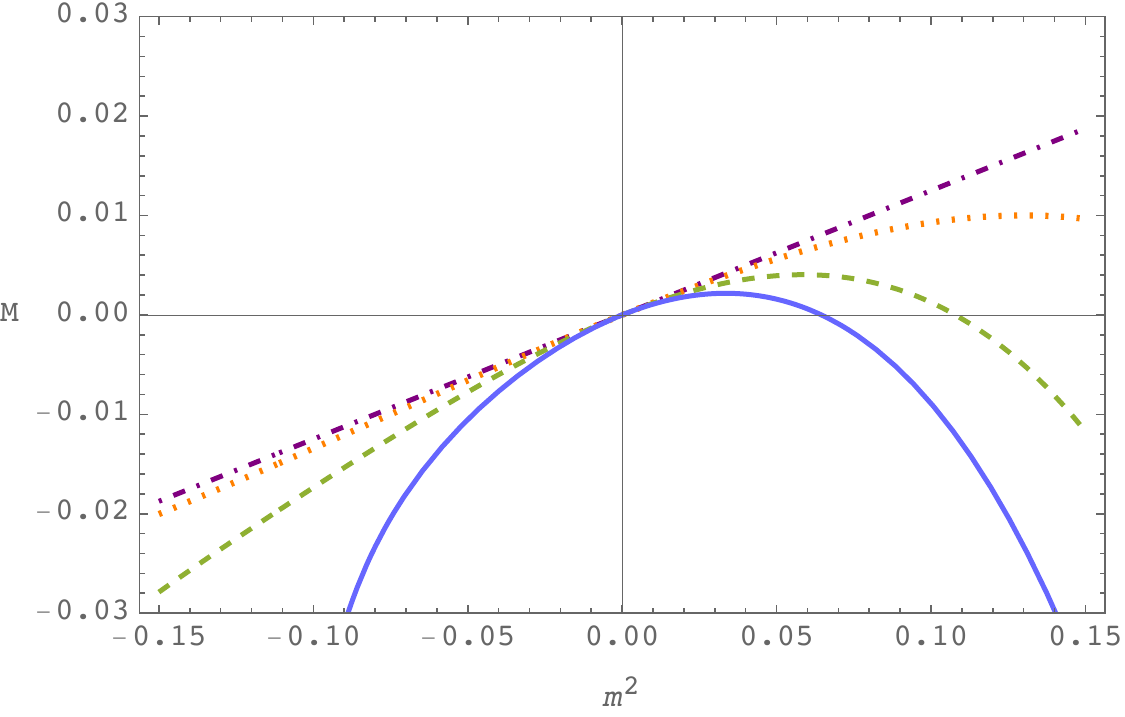} 
  \caption{Holographic mass of the accelerated BTZ pushed by a strut with respect to $m^{2}$~, with $\cA\ell = 0,1,2,3$ for the dot-dashed purple, dotted orange, dashed green, and solid blue curves, respectively.}
  \label{fig:massblm}
  \end{center}
\end{figure}

Finally, we can compute the energy of the theory with respect to $\partial_\tau$ by integrating the energy density of the fluid, thus
\begin{align}\label{M}
    M ={}& \int_{-\pi}^{\pi} d\phi \sqrt{-g_{(0)}}\langle T^\tau_{~\tau}\rangle 
 \nonumber \\ ={}& \frac{m^2\left[2\pi(2+m^2\cA^2\ell^2)-3m\cA^2\ell^2\sinh(2\pi m)\right]}{32\pi G \alpha}~,
\end{align}
of which the zero-accelerating limit gives \begin{align}
    \lim_{\cA\to0} M = \frac{m^2}{8G}~.
\end{align}
This precisely represents the BTZ mass normalised such that the pure vacuum energy corresponds to $m^2 \to -1$~. The behaviour of the mass can be seen in \autoref{fig:massblA} and \autoref{fig:massblm}.

\section{Euclidean Action: Counterterms and domain wall}\label{sec:HoloAction}

Holographic quantities are known to suffer from UV divergences. These are shown to be mapped to IR divergences appearing in the gravitational sector of theories on asymptotically anti-de Sitter spaces.
Consequently, defining observables requires having a well-defined renormalised action. This has been achieved in \cite{Henningson:1998gx, Balasubramanian:1999re, deHaro:2000vlm, Emparan:1999wa} by adding counterterms that depend only on intrinsic quantities of the boundary. For AdS$_3$ gravity, the renormalised Euclidean action consists of the Einstein--Hilbert action supplemented with the Gibbons--Hawking--York (GHY) term and the Balasubramanian--Kraus counterterm, ensuring a well-posed variational principle. However, the identifications performed in \autoref{sec:solutions} in order to construct the accelerating BTZ black hole, suggest that we need to identify the contributions coming from the $x=\rm const$ surface. Following \cite{Gregory:2001dn, Gregory:2001xu, Gregory:2001xu, Charmousis:2006pn}, the dynamics of the domain wall can be captured by considering the contributions of the Gibbons--Hawking terms associated with the surface (along each side) and the domain wall action \cite{Vilenkin:1984ib}, producing the Israel equation \eqref{Israel}. Then, the Euclidean action can be separated into two contributions 
\begin{align} \label{action}
    I_{\rm E} ={}& I_{\rm ren} + I_{\rm DW}~,
    \end{align}
    where 
    \begin{align}
    I_{\rm ren} ={}& -\frac { 1 }{ 16 \pi G } \int_{\mathcal{M}}{d^3x\sqrt {g }\left(R-2\Lambda\right)} \nonumber \\ {}&- \frac{1}{8\pi G} \int_{\partial \mathcal{M}} d^2 x \sqrt{h}\left(K- \frac{1}{\ell}\right)~, 
\end{align}
is the AdS$_{3}$ renormalised action \cite{Balasubramanian:1999re, Henningson:1998gx, deHaro:2000vlm}, and 
\begin{align}\label{DWaction}
    I_{\rm DW} = -\int_{\Sigma}d^2y\sqrt{\gamma}\left(\frac{1}{8\pi G}[{\cal K}]\rvert_-^+ + \sigma \right)~,
\end{align}
are the Gibbons--Hawking terms of the internal boundary and wall tension that give the correct equation of motion for the domain wall \eqref{Israel}.
Here $\Lambda = -\ell^{-2}$ is the cosmological constant, ${\cal M}$ corresponds to the bulk geometry restricted to some short IR regulator $\delta$ while $\partial {\cal M}$ is its boundary which is endowed with a metric $h_{ij}$ evaluated at $z = \delta$~. For the domain wall contribution we use
\begin{figure}
    \centering
\includegraphics{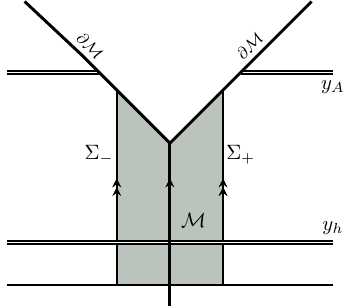}
    \caption{Class II solution with no accelerating horizon. The upper diagonal denotes the conformal boundary $\partial{\cal M}$~, and the lateral lines are the two internal boundaries $\Sigma_{-}$ and $\Sigma_{+}$ that are identified in order to obtain a compact horizon. The resulting spacetime corresponds to the accelerated BTZ solution with a domain wall extending from the black hole horizon to the deep interior.}
    \label{fig:Labeled}
\end{figure}
\begin{equation}
 \frac{1}{8\pi G}\int\limits_{\Sigma} d^2 x \sqrt{\gamma}\left[\mathcal{K}\right]\big\rvert_-^+ =- 2\int_{\Sigma} d^2 y \sqrt{\gamma}~\sigma~, 
\end{equation}
where the right-hand side is proportional to the action of the domain wall \cite{Vilenkin:1984ib}~, with $\sigma$ the tension computed in \eqref{tension}~. Thus, the contribution from the renomalised AdS$_{3}$ action is
\begin{align}
    I_{\rm ren} = \beta M - S - \frac{\beta\mathcal{A} m \sinh (\pi  m)}{4 \pi  \alpha  G}\left(\frac{1}{z_h}-\frac{1}{\delta}\right)~,
\end{align}
where $z_{h} : = r_h^{-1} + \cA \cosh(m\pi)$~. On the other hand, the domain wall gives
\begin{align}
    I_{\rm DW} = \frac{\beta\mathcal{A} m \sinh (\pi  m)}{4 \pi  \alpha  G}\left(\frac{1}{z_h}-\frac{1}{\delta}\right)~,
\end{align}
such that the total Euclidean action \eqref{action} yields the standard quantum statistical relation 
\begin{align}\label{IE}
    I_{\rm E } = I_{\rm ren} + I_{\rm DW} = \beta M - S~, 
\end{align}
where $M$ is the black hole energy found in \eqref{M}, $\beta$ and $S$ are the inverse of the temperature and the entropy found in \eqref{TS}, respectively. Note that if one considers only the Balasubramanian-Kraus counterterm on top of the Einstein--Hilbert and GHY terms, the resulting Euclidean action is divergent and the horizon contribution does not recover the black hole entropy. The domain wall extends from the interior (black hole horizon) to the conformal boundary contributing with a divergent term exactly as the one coming from the AdS$_{3}$ renormalised action but with an opposite sign, making the on-shell action well-defined and reproducing on-shell the quantum statistical relation between the gravitational Euclidean action and the Gibbs thermodynamic free energy.   

It is clear that this computation differs from the four-dimensional case \cite{Anabalon:2018qfv,Anabalon:2018ydc,Cassani:2021dwa}, where there is no explicit mention of the contribution of the cosmic string. In fact, this has been explained in the context of instantons in braneworld scenarios \cite{Gregory:2001dn}. The dynamics of the cosmic string, or \textit{vortex}, is introduced in the action as a codimension-2 energy density. On-shell, the extrinsic curvature contains an extra delta-2 function supported on the cosmic string, and therefore, the Gibbons--Hawking term cancels out the vortex. As a result, the renormalised Euclidean action in AdS$_4$ is enough to account for the thermodynamics of the accelerating black hole.


\section{Entanglement entropy}
In recent years, holographic entanglement entropy \cite{Ryu:2006bv,Ryu:2006ef} has been shown to be a fundamental ingredient in the understanding of the AdS/CFT correspondence. For example, it has served as a probe for quantum many-body systems, understanding black hole entropy, and the emergence of spacetime, see for instance \cite{Nishioka:2018khk} and reference therein. It is a powerful tool that provides valuable information on the dual field theory as it measures the degree of correlation between two subsystems. At the same time, it allows one to understand quantum information holographically using the nature of the bulk spacetime. The celebrated Ryu–-Takayanagi (RT) formula states that the holographic entanglement entropy can be obtained by minimising the area of a co-dimension two spatial\footnote{See \cite{Hubeny:2007xt} for the covariant generalisation.} hypersurface $\Gamma$ (referred to as the RT surface) whose boundary is anchored at the AdS conformal boundary, namely
\begin{align}\label{SERT}
    S_{\rm E} = \frac{A_\Gamma}{4G}~.
\end{align}
The RT surface divides the boundary into two subsystems $A$ and $A^c$ and \eqref{SERT} indicates the number of states on $A$ whose measures are consistent with the ones of $A^c$~.  
In general, the Euler--Lagrange problem is not easy to solve, and the profile of the RT surface is usually not obtained analytically as the existence of conservation laws is not guaranteed. Nonetheless, in three-dimensional gravity, as the theory lacks degrees of freedom; all solutions are locally AdS for which it is possible to reduce the complexity of the procedure by finding a good coordinate system. 

Following \cite{Johnson:2013dka}, we first consider a slice bounded by a line of latitude $\phi_0$ and treat $\phi$ as time in the minimisation problem. To obtain the RT surface for the accelerating BTZ, we use the mapping \eqref{Rmap} which allows us to find the extremal surface for the Rindler observer. The minimal surface is then parametrised by 
\begin{align}
    R_{\rm e}(\Theta) = \ell \left(1-\frac{\cosh^2\Theta}{\cosh^2\Theta_0}\right)^{-\frac12}~,
\end{align}
where $\Theta_0$ satisfies that the radial coordinates go to infinity at the endpoints. This is mapped to the boundary condition $r(\phi_0) = -(\cA \cosh(m\phi_0))^{-1}$~, such that the surface is anchored to the conformal boundary. Now we can map the surface to the coordinates used in \eqref{gtrphi}; for the sake of notation we define ${\cal B} := \cosh(m\phi_0)$ and $\phi_0 := m^{-1}\Theta_0$~, resulting in 
\begin{align}
    r_{\rm e}(\phi) =\frac{m \ell  \left(\alpha  \mathcal{A} m \ell  \cosh (m \phi )+{\cal B} \sqrt{{\cal B}^2-\alpha ^2 \sinh ^2(m \phi )-1}\right)}{\alpha  \left(\alpha^2\cosh ^2(m \phi )-{\cal B}^2\right)}~, 
\end{align}
whose expansion for small acceleration is
\begin{align}
    r_{\rm e}(\phi) ={}& \frac{m\ell}{\sqrt{1-\frac{\cosh^2(m\phi)}{\cosh^2(m\phi_0)}}}\nonumber \\ {}&- \frac{m^2\cA\ell\cosh(m\phi) }{\cosh^2(m\phi_0) - \cosh^2(m\phi)}+{\cal O}(\cA^2)~.
\end{align}
Substituting the parametrisation into the area functional, one obtains the value of the minimal area that is proportional to the holographic entanglement entropy. Despite the simplicity of the last expression, obtaining the area is quite involved. It is divergent at $\phi = \phi_0$~, and therefore a short distance cutoff $\epsilon$ must be introduced. Then, following \cite{Ryu:2006bv}, we consider the integration from $\epsilon$ to $\phi_0 - \epsilon$~ and expanding again for small acceleration, we get that the entanglement entropy \eqref{SERT} becomes
\begin{align} \label{SE}
    S_{\rm E} ={}&\frac{c}{3}\log\left[\frac{\beta}{\pi\epsilon}\sinh\left(\frac{\pi L}{\beta}\right)\right]\nonumber \\ {}&-2\cA\ell^2\left({\frac{2\pi\ell}{\beta\epsilon}\sinh\left(\frac{\pi L}{\beta}\right)}\right)^{\frac12}\tanh\left(\frac{\pi L}{2\beta}\right) \nonumber \\ {}&-\frac{\cA^2\ell^4\pi}{\beta\epsilon}\sinh\left(\frac{\pi L}{\beta} \right)-\dots~,
\end{align}
where we have rewritten $L := 4\ell \phi_0$ to relate it with the length of the entangling region. Note that, as the temperature is independent of the acceleration, when mapping to Rindler AdS the leading order corresponds to the usual result for the BTZ black hole \cite{Ryu:2006bv}, However, the next to leading order gives subleading divergences which decrease the amount of entanglement with the acceleration growth.
In fact, from the perspective of the black hole solution, as the acceleration -- or in other words, the conical deficit -- increases, the size of the boundary region that is accessible decreases, as can be seen from \autoref{fig:c2const}. Therefore, we can interpret the subleading behaviour of the entanglement as an indication of some information loss due to acceleration. 

\section{Discussion}

In this work, we have described the boundary spacetime associated with accelerating black holes in 2+1 dimensions. Our starting point has been a concise and pedagogical review of the construction of the three-dimensional accelerating geometries presented in \cite{Astorino:2011mw,Xu:2011vp,Arenas-Henriquez:2022www}. We have analysed the origin of the acceleration in three dimensions and the proper identifications of the geometry that allows for black hole interpretation. Particular emphasis has been given to the case in which the spacetime represents an accelerating BTZ black hole pushed by a strut.
This case is particularly appealing, as it allows for an accelerating horizon, however, for thermodynamic purposes, we focussed on the case where there is just a black hole horizon.

Since the conformal boundary is defined by a surface that is parameterised as $r=r(\phi)$~, determining the boundary metric becomes a non-trivial task. To address this, we introduced an alternative coordinate system that incorporates a new ``holographic coordinate'', $z$~, that is normal to the boundary \cite{Hubeny:2009kz, Cassani:2021dwa}. In this framework, the metric can be expressed in terms of a radial ADM foliation, revealing crossed terms that are typically suppressed when writing the C-metric in the FG fashion \cite{Anabalon:2018ydc,Anabalon:2018qfv}. In fact, as noticed in \cite{Hubeny:2009kz}, beyond the $z$ coordinate can be identified with the FG coordinate only up to the leading order. Nonetheless, the first order of the expansion fully determines the variational principle and therefore, the structure of the boundary stress tensor.  We obtained the black hole mass by mapping it to the energy of the dual CFT and verified that it recovers the BTZ mass in the zero acceleration limit. Additionally, we formulated the holographic stress tensor using the fluid/gravity correspondence, wherein the dual CFT is interpreted as a perfect fluid with non-constant pressure on a curved background. This is in contrast to the four-dimensional case, which incorporates shears and corrections arising from the non-conformal flatness of the boundary metric.

Next, we computed the renormalised action by employing the standard counterterm prescription in AdS/CFT, as developed in \cite{Henningson:1998gx, deHaro:2000vlm, Emparan:1999fd}. We found that the on-shell action gives the quantum statistical relation which relates the partition function and the Gibbs free energy as expected but with two additional terms.
These terms, in principle, contain an extra divergence that comes from the boundary terms of the gravitational action (Gibbons--Hawking--York term and Balasubramanian--Krauss counterterm) and a finite contribution of the black hole horizon. In fact, due to the construction of the accelerating BTZ black hole, it is necessary to include contributions from the internal boundaries which are on the same footing as the GHY term. These terms ensure a well-posed variational problem producing the Israel junction conditions that govern the dynamics of the domain wall. Making use of the Israel equations explicitly, it is possible to trade the extrinsic curvature for the energy density of the wall and therefore, express the additional term as the Nambu--Goto action of the domain wall. In a similar spirit to  \cite{Martinez:1990sd, Eune:2013qs, Arias:2019zug}, the higher-codimension defect induces extra contributions into the partition function modifying the thermodynamics of the system under consideration. Geometrically, the domain wall extends from the deep interior to the boundary of the spacetime generating a divergence at $z=0$~. Therefore, its contribution to the total Euclidean action must be considered in order to obtain the correct quantum statistical relation.

We closed our study by considering the mapping between the accelerated BTZ black hole and the Rindler observer which allows us to obtain the Ryu--Takayanagui surface and to compute the holographic entanglement entropy on the dual CFT. 
We found that the well-known logarithmic divergence of the entanglement entropy in a thermal conformal field theory holds in this context. However, we also discovered new subleading divergences that are proportional to acceleration and possess a negative sign.  
In \cite{Arenas-Henriquez:2022www}, it is shown that the boundary region of the spacetime is altered by the tension of the domain wall. From \autoref{fig:c2const}, it is clear that the access to the boundary depends on the value of the acceleration. In fact, the behaviour of the entanglement is consistent with this interplay between acceleration and boundary: as the acceleration increases, a bigger portion of the AdS boundary is cut out and therefore, there is information that is lost in the dual field theory as suggested by \eqref{SE}. As far as the authors' knowledge, such subleading behaviour has not been observed before in the literature. 
It is also important to note that the procedure is specific to three dimensions, as only massless four-dimensional accelerating solutions can be mapped to the Rindler patch and the identification of the RT surfaces becomes a highly non-trivial task. This realisation highlights that three dimensions offer a unique yet comprehensive setting for exploring holographic two-dimensional CFTs in the presence of acceleration.

In the future, an important aspect that requires further investigation is the establishment of a consistent thermodynamic description of these black holes. This entails studying the first law, Smarr relation, isoperimetric inequality \cite{Cvetic:2010jb} and the whole machinery of black hole thermodynamics. In fact, given the complexity of the mass \eqref{M} and entropy \eqref{TS}, verifying whether these black holes adhere to the first law is not straightforward. Upon a simple variation of these quantities, it becomes apparent that there exists a possibility that, unlike slow-accelerating black holes in AdS$_{4}$~, the system might not conform to a first law and thus may not be in thermal equilibrium. Nonetheless, this is not yet clear, as there are several issues that require consideration before making such a statement. Given that we have obtained the quantum statistical relation, it seems very plausible to have a full Euclidean thermodynamic prescription for accelerating black holes in 2+1 dimensions as it has been done for the four-dimensional counterpart in \cite{Anabalon:2018ydc}.  
Additionally, it would be intriguing to investigate the role of acceleration in the dual theory using the extended first law developed in \cite{Appels:2017xoe,Gregory:2020mmi}. This modified first law incorporates new chemical potentials that are conjugate to the cosmic string tension, potentially providing insights into the physical properties associated with these additional chemical potentials. Furthermore, recent work \cite{Visser:2021eqk, Cong:2021fnf, Ahmed:2023snm} has shown that the extended first law of black hole thermodynamics introduces a new chemical potential responsible for the change in the central charge of the dual CFT. It would be interesting to see how the domain wall tension plays a role in the first law of thermodynamics of the boundary theory. 

Another interesting direction that would shed light on the role of acceleration from the dual CFT perspective is to explore the hydrodynamic behaviour of the holographic stress tensor for four-dimensional accelerating black holes. While the stress tensor has been expressed within the framework of fluid/gravity correspondence \cite{Anabalon:2018ydc}, it remains unclear whether it possesses a valid hydrodynamic description that allows for the identification of associated transport coefficients. It would be interesting to see whether acceleration plays a significant role in determining the transport coefficients and if they can be utilised to describe more realistic field theories. Additionally, an expansion regarding the fluid velocity and acceleration of the dual fluid stress tensor is still an open question. The three-dimensional case studied in this paper serves as a good starting point, as the solution is relatively simple yet rich enough to generate a stress tensor that exhibits non-constant pressure. This enriches the opportunities for studying more realistic systems through the scope of fluid/gravity correspondence.

\section*{Aknowledgments}
We thank Moh Al Attar, Jose Barrientos, Saghar Hosseini, Viktor Matyas, Olivera Miskovic, and Rodrigo Olea for helpful discussions. {\sc Gah} would like to thank Aristomenis Donos for his invaluable comments and insightful discussions, which contributed to the improvement of this paper. {\sc Gah} also thanks Mohamed Anber and Jerome Gauntlett for their useful comments and discussions regarding this work. {\sc Ac} and {\sc Fd} would like to thank the hospitality of the Institute of Mathematics of the Czech Academy of Science during the final stage of this project. The work of {\sc Gah} is funded by {\sc Becas Chile} (ANID) Scholarship N$^{\rm o}~ 72200271$~. {\sc Ac} work is partially funded by FONDECYT Regular grant No. 1210500 and Primus grant PRIMUS/23/SCI/005 from Charles University.
The work of {\sc Fd}~is supported by {\sc Beca Doctorado nacional} 
(ANID) 2021 Scholarship N$^{\rm o}~21211335$~, ANID/ACT210100 Anillo Grant ``{\sc Holography and its applications to High Energy Physics, Quantum Gravity and Condense Matter Systems}'' and FONDECYT Regular grant No. 1210500.  
The work of {\sc Rg} is supported in part by the STFC Consolidated Grant ST/P000371/1. {\sc Rg} would like to thank the Perimeter Institute for Theoretical Physics for hospitality.
Research at Perimeter Institute is supported by 
the Government of Canada through the Department of Innovation, 
Science and Economic Development Canada and by the Province of 
Ontario through the Ministry of Colleges and Universities.

\begin{appendix}
\section{FG expansion}\label{App:FG}
In \cite{Anabalon:2018qfv} it was shown that the metric can be cast in a FG gauge 
\begin{align} \label{fg}
ds^{2} ={}& \frac{\ell^{2}}{z^{2}}dz^{2}+\frac{\ell^2}{z^{2}}\left(g_{(0) ij} +  \dots \right.\nonumber \\ {}&\left.+ z^{d}\left(g_{(d)ij}+h_{(d)ij}\log (z)\right) + \dots \right)dx^{i}dx^{j}~,  
\end{align}
which in three dimensions, the expansion terminates at order $z^4$~, and we do not consider the logarithmic contribution explicitly.  
Applying the same coordinate transformation of \cite{Arenas-Henriquez:2022www} 
\begin{eqnarray}\label{coordFG}
y={}&-\frac{1}{A r}=\cos\xi+\sum\limits_{m=1}^{\infty} F_{m}(\xi)z^{m}~, \nonumber \\
x ={}& \cos \xi +\sum\limits_{m=1}^{\infty}G_{m}(\xi)z^{m}~,
\end{eqnarray}
such that, the new coordinate $\xi$ is now perpendicular to the (conformal) boundary of the spacetime. The functions $F_m(\xi)$ and $G_m(\xi)$ are fixed by requiring the fall of conditions of \eqref{fg} such that there are no crossed terms $g_{z i}$~. The coefficients can be solved order by order completely, up to $F_1(\xi)$ which cannot be fixed and appears as a conformal factor of the boundary metric $g_{(0)}$~, which is consistent with the fact that the conformal boundary of AdS does not correspond to a fixed metric but to conformal equivalence classes. As explained in \autoref{sec:HoloAction}, in three dimensions, besides the Weyl anomaly, the holographic quantities are not conformal invariant and a different coordinate system is needed in order to compute on-shell. 
Considering the expansion for the accelerated BTZ black hole (although the claims and calculation of this appendix hold for the other black hole solutions as well), one gets \cite{Arenas-Henriquez:2022www}
\begin{align}\label{g0fg}
    ds_{(0)}^2 = \omega^2\left(-d\bar{t}^2 + d\xi^2\right)~,
\end{align}
where $\bar{t} = \alpha\ell t$ and $\omega = \omega(\xi)$ is an arbitrary function which determine different conformal representatives. The boundary curvature reads
\begin{align}\label{Rg0fg}
    {\cal R}[g_{(0)}] ={}& \frac{2\Upsilon}{\ell^2\omega(\xi)}\Bigg[\Upsilon\sin^2\xi\left(\frac{\omega''}{\omega}-\frac{\omega'^2}{\omega^2}\right) \nonumber \\ {}&+\cos\xi\left(1-3A^2\ell^2\sin^2\xi\right)\frac{\omega'}{\omega}\Bigg]~,
\end{align}
where $\Upsilon = 1-A^2\ell^2\sin^2\xi$~. 
Other quantities such as the stress tensor and Weyl anomaly have been computed with this method in \cite{Arenas-Henriquez:2022www}. But as aforementioned, only the Weyl anomaly is conformal invariant and, indeed, gives the Brown--Henneaux central charge for any representative. Other features are also independent of $\omega$~, such as that the holographic stress tensor is covariantly constant with respect to $g_{(0)}$ and its transformation properties. Nonetheless, the two-dimensional stress tensor is a quasi-primary operator and to compute conserved quantities, such as the energy, depends on the Schwarzian derivative of it. In \cite{Arenas-Henriquez:2022www} it has been chosen $\omega(\xi) = 1$~, which renders the boundary metric to be flat, as can be seen from \eqref{Rg0fg}. Comparing with the boundary metric found in \eqref{g0metric}, taking $\omega(\xi) = G(\xi)$ in \eqref{Gomega} equals both backgrounds. Therefore, we shall not compare the energy found via holographic methods.  

Finally, the holographic stress tensor can be cast in the same fashion as in \eqref{fluidT} with velocity \begin{align}
    u_i = \omega(\xi)\delta_i^t~,
\end{align}
pressure
\begin{align}
    p = \frac{(1+ m^2 \cA^2\ell^2)\omega^2+(\xi^2-1)\Upsilon^2\omega'^2}{16\pi G \ell \alpha^2\omega^4}~,
\end{align}
and energy density 
\begin{align}
 \rho ={}& \frac{m}{16 \pi  \alpha  G \Upsilon  \omega (\xi )^2\sin\xi} \Bigg[  2 \Upsilon  \omega (\xi ) \times \nonumber \\ {}&\bigg( \omega '(\xi ) \left(3 m^2\mathcal{A}^2\ell^2 \sin^2\xi+1\right)\cos\xi - \Upsilon  \omega ''(\xi )\sin^2\xi\bigg)  \nonumber \\ & 
+ \omega (\xi )^2 \left(m^2\mathcal{A}^2\ell^2+1\right)+  3\Upsilon ^2 \omega '(\xi )^2\sin^2\xi\Bigg]~.
\end{align}

Finally, we can now compute the Euclidean action with the standard counterterms coming from holographic renormalisation with no need to include the domain wall action as the metric is explicitly in the FG gauge. For the accelerated BTZ pulled by a wall one gets that for the action to become finite, one must consider the extra counterterm that accounts for the extra logarithmic divergence that appears in odd dimensions \cite{Henningson:1998gx, deHaro:2000vlm, Emparan:1999wa} of the form
\begin{align}
    I_{\rm log} = \log \delta\int dt d\xi \sqrt{g_{(0)}}
    {\rm tr}~g_{(2)}~,
\end{align}
where $\delta$ is the IR regulator, and traces are taking respect to the boundary metric $g_{(0)}$~. However, in $d=2$ boundary dimensions, this term fails to furnish any contribution to the holographic stress tensor since this term is proportional to the boundary curvature in the renormalised action, which is a topological invariant in two dimensions \cite{deHaro:2000vlm}. Then, the Euclidean action becomes finite, but one is not able to identify the thermodynamic quantities as integration is not possible for an arbitrary $\omega(\xi)$~. The finite value of the action is rather lengthy and we shall not present it here. 

\section{Results for class I$_{\rm C}$}\label{App:Ic}

Generically, Class I describes the geometry of accelerating particle-like solutions. Nevertheless, a particular case dubbed Class I$_{\rm C}$~, represents an accelerating black hole solution with no continuous limit to the standard BTZ geometry. The novel solution was found in \cite{Arenas-Henriquez:2022www} by considering Class I geometries in a rapid phase, $A^2\ell^2 \geq 1$~, in which there is a Killing horizon at $y_h = \sqrt{1-A^{-2}\ell^{-2}}$~. Then, following the procedure of \autoref{sec:solutions} to include a domain wall at some $x_{\rm max}$~, with tension
\begin{align}
    \sigma = \frac{A}{4\pi G}\sqrt{1-x_{\rm max}^2}~.
\end{align}
Then, using the coordinate transformation
\begin{align}
    t = \frac{{\cal A}m^2}{\alpha}\tau~,\qquad y = \frac{1}{{\cal A}r}~,\qquad x = \cos(m\phi)~,
\end{align}
renders \eqref{eq:metricxy2} (for Class I) into
\begin{align}
    ds^2 = \frac{1}{\Omega^2}\left(-f(r)\frac{d\tau^2}{\alpha^2}+\frac{dr^2}{f(r)} + r^2d\phi^2\right)~,
\end{align}
where
\begin{align}
    f(r) ={}& \frac{r^2}{\ell^2} - m^2(\cA^2r^2-1)~,\nonumber \\ \Omega ={}& \cA r \cos(m\phi)-1~,
\end{align}
and the tension 
\begin{align}
    \sigma = \frac{{\cal A}m}{4\pi G}\sin(m\pi)~.
\end{align}

As before, the conical deficit is regulated by the parameter $m$~, which relates with the upper bound of the $x$ coordinate as
\begin{align}
    x_{\rm max} = \cos(m\pi)~.
\end{align}
As explained in \cite{Arenas-Henriquez:2022www}, $x_{\rm max} \in (y_h,1)$~, with $y_h \geq 0$ in order to have a single compact horizon. This implies that there is a maximum value for the mass parameter as  
\begin{align}
    m < \frac{\arccos(y_h)}{\pi}~,
\end{align}
and a minimum value $m > 0$~. Otherwise, the solution would present a non-compact horizon. 
Moreover, as now the mass parameter $m$ is bounded, there is also a constraint in the possible values of the acceleration in order that geometry exists, given by 
\begin{align}\label{IcConst}
   \frac{1}{m} \leq {\cal A}\ell < \frac{1}{m\sin(m\pi)}~. 
\end{align}
This implies that ${\cal A}{\ell} \geq 2$ in order to have a black hole, showing explicitly that the zero-acceleration limit is not well-defined, and the solution is not continuously linked with the BTZ black hole. 
The thermodynamic properties of the horizon, namely, horizon
radius, Hawking temperature and entropy are
\begin{align}\label{TSIc}
    r_h ={}& \frac{m\ell}{\sqrt{m^2\cA^2\ell^2-1}}~,\\ \nonumber T ={}& \frac{1}{\beta} = \frac{|f'(r_h)|}{4\pi\alpha} = \frac{\sqrt{m^2\cA^2\ell^2-1}}{2\pi\ell\alpha}~,\\ S ={}& \frac{\ell}{G} \arctanh\left[\left( \sqrt{m^2\cA^2\ell^2-1}+m\cA\ell\right)\tanh\left(\frac{m\pi}{2}\right)\right]~,\nonumber \\ 
    \alpha ={}& \sqrt{m^2\cA^2\ell^2-1}~.\nonumber 
\end{align}
Finally, $\alpha$ is the constant used in the transformation \eqref{Rmap} to identify the Rindler proper time. 

\begin{figure}
\centering
\includegraphics[scale=0.45]{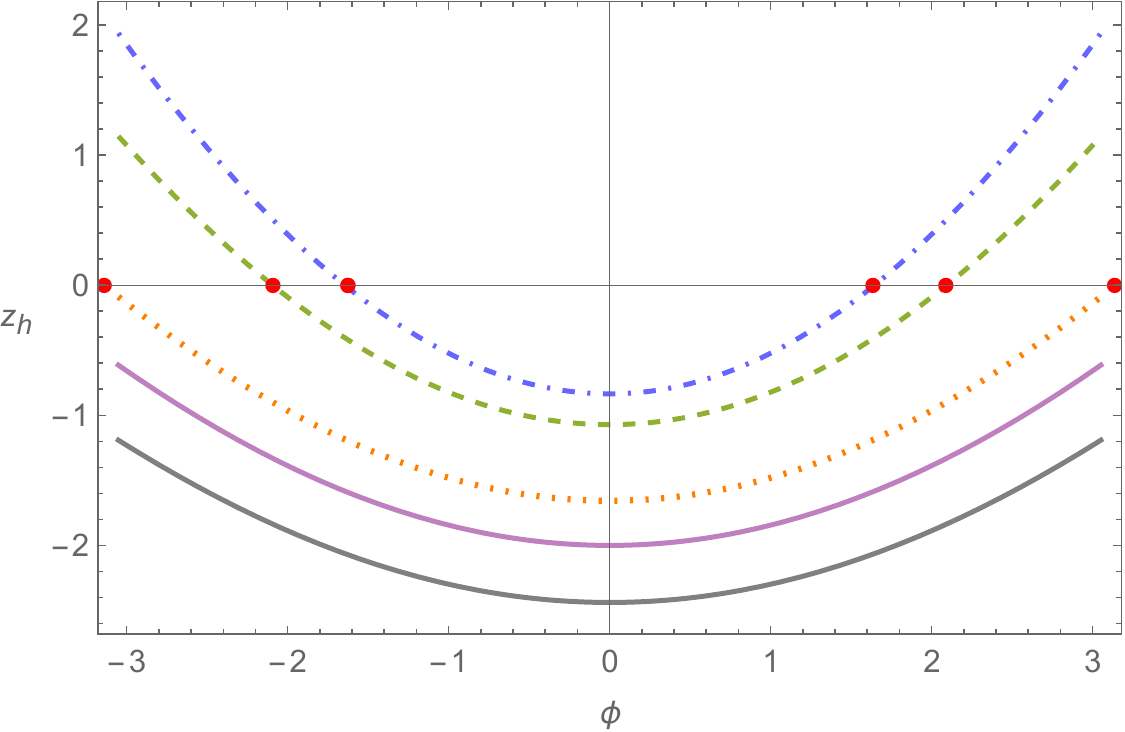} 
  \caption{Horizon radius with $\ell = 1$ and $m = 0.25$~. The grey and purple solid lines correspond to $\cA = 4.5, 5$~, respectively, and both values are admissible for the horizon. Red dots indicate points where the horizons intersect with the conformal boundary. The dotted orange curve corresponds to the critical value $\cA \ell = (m\sinh m\pi)^{-1} = 5.656$ where the horizons touches the conformal boundary, $z=0$~, exactly at the endpoints $\phi = \pm \pi$~. The dashed green and dot-dashed blue correspond to $\cA = 8, 10$~, respectively, where both cases break the inequality \eqref{IcConst}.  }
  \label{fig:horzclassIc}

\end{figure}
In order to compute the mass, we apply the same ideas as in \autoref{sec:HoloAction}. Firslty, we introduce 
\begin{equation}
    \frac{1}{r} = z + \cA \cos(m\phi)~,
\end{equation}
such that the conformal boundary is now located at $z = 0$~. This implies that the horizon is no longer located on a constant surface. Nonetheless, the inequality \eqref{IcConst} ensures that the horizon does not intersect with the conformal boundary, as can be seen from \autoref{fig:horzclassIc}.

Using \eqref{Tmunu} to compute the holographic stress tensor, one obtains 
\begin{align}
    \langle T^\tau_{~\tau} \rangle ={}& \frac{m^2\ell}{32\pi G}\left(2-m^2\cA^2\ell^2 + 3m^2\cA^2\ell^2\cos(2m\phi)\right)~, \nonumber \\ \langle T^\phi_{~\phi} \rangle ={}& \frac{m^2\ell}{16\pi G}\left(m^2\cA^2\ell^2 \cos^2(m\phi)-1\right)~, 
\end{align}
whose trace recovers exactly \eqref{weylanomaly}, with $c = 3\ell/2G$ the Brown--Henneaux central charge. Furthermore, the boundary stress tensor is covariantly conserved with respect to the boundary metric $g_{(0)}$~, and can also be written as the one of a perfect fluid \eqref{fluidT} with non-constant pressure and density. 
Moreover, using \eqref{M}, one gets
\begin{align}\label{MIc}
M = \frac{m^2\left[2\pi m^2\cA^2\ell^2 - 3m\cA^2\ell^2\sin(2\pi m)-4\pi\right]}{32\pi G \alpha}~,
\end{align}
corresponding to the I$_{\rm C}$ black hole mass. 

Finally, the Euclidean action shares the same properties as the Class II black holes; the Balasubramanian--Krauss action has an extra divergence and an extra horizon finite contribution due to the acceleration that is removed by the inclusion of the domain wall action \eqref{DWaction}. Then, its on-shell value 
\begin{align}
    I_{\rm E} = \beta M - S~,
\end{align}
satisfies the standard quantum statistical relation with the Gibbs free energy.

\vspace{3cm}
\end{appendix}
\bibliographystyle{JHEP}
\bibliography{ref}
\end{document}